# Electrical detection in two-terminal perpendicularly magnetized devices via geometric anomalous Nernst effect

Jiuming Liu[1†], Bin Rong[1†], Hua Bai[2†], Xinqi Liu[3,4], Yanghui Liu[1], Yifan Zhang[1], Yujie Xiao[4], Yuzhen Liang[1], Qi Yao[3,4], Liyang Liao[5], Yumeng Yang[1], Cheng Song[2*], and Xufeng Kou[1,3*]

[1]School of Information Science and Technology, ShanghaiTech University, Shanghai 201210, China

[2]Key Lab Advanced Materials (MOE), School of Materials Science and Engineering, Beijing Innovation Center for Future Chips, Tsinghua University, Beijing 100084 China

[3]ShanghaiTech Laboratory for Topological Physics, ShanghaiTech University, Shanghai 201210, China

[4]School of Physical Science and Technology, ShanghaiTech University, Shanghai 201210, China

[5]Institute for Solid State Physics, University of Tokyo, Kashiwa 277-8581, Japan

†These authors contributed equally to this work.

*Correspondence to: kouxf@shanghaitech.edu.cn; songcheng@mail.tsinghua.edu.cn

**KEYWORDS:** Anomalous Nernst effect, device geometry, spin-orbit coupling, two-terminal spin-orbit-toque memory, nonlinear effects

**The non-uniform current distribution arisen from either current crowding effect or hot spot effect provides a method to tailor the interaction between thermal gradient and electron transport in magnetically ordered systems. Here we apply the device structural engineering to realize an in-plane inhomogeneous temperature distribution within the conduction channel, and the resulting geometric anomalous Nernst effect (GANE) gives rise to a non-zero 2nd-harmonic resistance whose polarity corresponds to the out-of-plane magnetization of Co/Pt multi-layer thin film, and its amplitude is linearly proportional to the applied current. By optimizing the aspect ratio of convex-shaped device, the effective temperature gradient can reach up to 0.3 K/μm along the *y*-direction, leading to a GANE signal of 28.3 μV. Moreover, we demonstrate electrical write and read operations in the perpendicularly-magnetized Co/Pt-based spin-orbit torque device with a simple two-terminal structure. Our results unveil a new pathway to utilize thermoelectric effects for constructing high-density magnetic memories.**

**Introduction**

Efficient electrical detection of the magnetic moment is essential for advancing spintronic logic memory and sensor applications[1-4]. Based on the critical role of spin-related scatterings on the electron charge transport, various magneto-electric effects have been applied to probe the magnetic information of the host device. For instance, anisotropic magnetoresistance (AMR), which stems from the spin-dependent conductivity, provides the magnetic field direction by tracking the resistance change with respect to the rotation angle between the magnetic moment and the spin-polarized current[5,6]; the spin-dependent scattering-induced anomalous Hall effect (AHE) can directly reflect the change of magnetization orientation under various external stimuli[7,8]; and giant magnetoresistance (GMR) is used to represent the parallel and antiparallel configurations of the magnetic moments in stacked ferromagnetic layers[9,10].

In align with the above 1st-order magneto-transport phenomena, relevant 2nd-harmonic non-reciprocal electrical responses such as unidirectional spin Hall magnetoresistance (USMR) and anomalous Nernst effect (ANE) have also been identified as effective approaches to readout magnetic-related information[11-14]. With the breaking of time-reversal symmetry by the magnetic moment *m*, the USMR/ANE-associated non-reciprocal charge flow could give rise to a unique 2nd-harmonic resistance with a one-fold symmetry in reference to magnetization, hence enabling the differentiation of magnetic moments with opposite polarizations[11,15]. In contrast to USMR which only allows for the in-plane magnetization detection (*i.e.*, because the spin accumulation at the ferromagnet/heavy metal interface points towards the *y*-axis when the charge current flows along the *x*-direction)[16,17], ANE-related effects, in principle, are able to interplay with both perpendicular and in-plane magnetic moments. As illustrated in Figure 1a, the presence of a temperature gradient ($\nabla T$) in a magnetic system can generate an electric field $\boldsymbol{E}_{\text{ANE}} = S \cdot \boldsymbol{m_0} \times \nabla T$, where $S$ is ANE coefficient, and $\boldsymbol{m_0}$ is the unit vector of $\boldsymbol{m}$[18-20]. Considering that conventional conductive magnetic thin films are normally grown on insulating substrates, the Joule heating from the applied AC current ($I = I_0 \cdot \sin\omega t \cdot \hat{\boldsymbol{x}}$) would cause a $z$-direction temperature gradient, and the ANE-induced 2nd-harmonic resistance $R_{2\omega}^{\text{ANE}}$ can henceforth be probed between the Source (S) and Drain (D) terminals if the sample is magnetized along the $y$-direction[21-23]. Following the same operating principle, the two-terminal magnetization detection of a perpendicular magnetic anisotropy (PMA) system (*i.e.*, $\boldsymbol{m} = m_z \cdot \hat{\boldsymbol{z}}$) is feasible when the temperature gradient is along the $y$-axis, yet how to generate such a local $\nabla_y T$ within a two-terminal device remains to be explored.

In this article, we propose and demonstrate the geometric anomalous Nernst effect (GANE), which enables the electrical detection of the out-of-plane magnetization in Co/Pt multilayers. The non-uniform current distribution in the double-cross device introduces a lateral temperature gradient along the conduction channel, which in turn contributes to 2nd-harmonic magneto-resistance loops with opposite polarities among different contact pairs. Guided by the COMSOL simulations, we optimize the structural aspect ratio of the convex-shaped device to enhance the GANE signal via the hot spot effect. By further incorporating the GANE scheme into the two-terminal spin-orbit torque (SOT) device, electrical write and read operations are demonstrated at room temperature.

**Results**

**Demonstration of geometric anomalous Nernst effect.** In general, an in-plane temperature gradient can be aroused by either the current crowding effect or the hot spot effect. Specifically, when electrons flow through a non-uniform path, the current continuity law dictates that the narrower region with a higher current density would always accumulate a more pronounced Joule heating[24,25]. In this context, the relevant steady-state spatial temperature mapping would display a hot-cold-hot configuration in the cross-bar structure, as shown in the upper panel of Figure 1b. Meanwhile, the Neumann boundary condition (*i.e.*, which states that the electrical field $\boldsymbol{E}$ has to be tangent to the sample surface[26]) invariably leads to a higher current density distribution around the deformation corners of the device structure, and the resulting hot spots would introduce a temperature gradient along the $y$-axis in the convex-shaped device (lower panel of Figure 1b). As a result, if such device structures are further incorporated into the PMA systems, the geometry-induced in-plane temperature gradient $\nabla T = A \cdot \langle I^2 \rangle \cdot R \cdot t_0$ (*i.e.*, where $A$ is the characteristic heat-to-temperature gradient conversion coefficient, $R$ is the channel resistance, and $t_0$ is the time period for the system to reach the thermal steady state) hence gives rise to a non-zero $\boldsymbol{E}_{\text{ANE}}$ field whose 2nd-harmonic component $E_{2\omega}^{\text{GANE}} = \frac{1}{2} S \cdot A \cdot R \cdot t_0 \cdot I_0^2 \cdot \sin(2\omega t)$ enables the electrical readout of the out-of-plane magnetization by means of the 2nd-harmonic magneto-resistance signals (*i.e.*, dubbed as GANE). In the following section, we will demonstrate the operating principle of GANE and apply it for device applications.

Experimentally, the Co/Pt multi-layer thin film that consists of 8×[Pt(0.8 nm)/Co(0.4 nm)] and a 2nm Pt buffer layer was deposited on the MgO(111) substrate by magnetron sputtering at room temperature (Figure 1c)[27]. After sample growth, standard nano-fabrication process was used to fabricate devices on the as-grown sample, and the temperature distributions within the specific device structures were simulated by COMSOL Multiphysics Simulation Software (Supplementary Information S1). As exemplified in Figure 2a, in a double-cross device with each arm size of 120 μm (*L*) × 30(20) μm (*W*), the applied current of |$I_0$| = 40 mA is able to introduce the expected hot-cold-hot mapping along the conduction channel due to the current crowding effect, and the temperature difference is as large as Δ*T* = 15 K to ensure a sizable GANE signal (*i.e.*, the non-uniform current distribution mappings are provided in Supplementary Information S2).

Inspired by these simulation results, magneto-transport measurements were subsequently performed on the fabricated device where both first and second harmonic resistances between the six probing contacts (*i.e.*, denoted as $R_{\omega/2\omega}^{xx,6-4}$, $R_{\omega/2\omega}^{xx,1-3}$, $R_{\omega/2\omega}^{xy,2-5}$, $R_{\omega/2\omega}^{xy,3-4}$, and $R_{\omega/2\omega}^{xy,1-6}$ respectively) were simultaneously recorded under a fixed AC current (*i.e.*, whose amplitude was kept as $I_0$ = 40 mA, and the lock-in frequency was set as 177.78 Hz) and a varied perpendicular magnetic field ranging from −400 mT to +400 mT. As displayed in Figures 2c-g, the two longitudinal resistances ($R_{\omega}^{xx,1-3}$ and $R_{\omega}^{xx,6-4}$) remain almost constant against the magnetic field variation, indicating a negligible magneto-resistance ratio of the Co/Pt system. In the meanwhile, three identical square-shaped anomalous Hall resistance loops with the coercivity field of $H_C$ = 200 mT are obtained between the 2-5 ($R_{\omega}^{xy,2-5}$), 3-4 ($R_{\omega}^{xy,3-4}$), and 1-6 ($R_{\omega}^{xy,1-6}$) contacts, which attest the robust perpendicular magnetic anisotropy in our sample. On the other hand, the second harmonic components of these resistances all exhibit magnetic field-dependent hysteresis loops, yet their polarities depend on the contact locations. In particular, according to the simulated heatmap of Figure 2a, the "cold zone" in the center area produces a negative temperature gradient on the left part of the Co/Pt-based double-cross device. Under such circumstances, the GANE-induced transverse electric field would point to the +y-direction when the magnetic moment of Co is aligned along the +z-axis, therefore leading to a negative $R_{2\omega}^{xy,1-6} = (V_{2\omega}^1 - V_{2\omega}^6)/I_0 < 0$ state when a large positive magnetic field is applied (lower panel of Figure 2f). On the contrary, the presence of $\nabla_x T > 0$ on the right side of the conduction channel brings about $\boldsymbol{E}_{\text{ANE}} = -|E_{\text{ANE}}| \cdot \hat{y}$, which is responsible for the counter-clockwise $R_{2\omega}^{xy,3-4}$- $\boldsymbol{B}$ curve observed in Figure 2e. More importantly, this geometric anomalous Nernst effect also re-distributes the electric potential within the double-cross device, namely the 2$^{\text{nd}}$-harmonic voltages measured at the four probing contacts are $V_{2\omega}^1 = V_{2\omega}^4 = -10$ μV and $V_{2\omega}^3 = V_{2\omega}^6 = +10$ μV in reference to the input current level of $I_0$ = 40 mA and $B_z$ = +400 mT (*i.e.*, the zero potential was defined as the white dashed line in Figure 2a). Consequently, the corresponding 2$^{\text{nd}}$-harmonic longitudinal resistances are $R_{2\omega}^{xx,1-3} = (V_{2\omega}^1 - V_{2\omega}^3)/I_0 = -0.5$ mΩ and $R_{2\omega}^{xx,6-4} = (V_{2\omega}^6 - V_{2\omega}^4)/I_0 = +0.5$ mΩ, as experimentally justified in Figures 2b-c. Besides, owning to a rather uniform thermal distribution between the 2$^{\text{nd}}$ and 5$^{\text{th}}$ probing contacts, it is noted that a nearly diminished $R_{2\omega}^{xy,2-5}$ signal was captured during the magnetic field sweeping, which further validates the GANE-induced 2$^{\text{nd}}$-harmonic magneto-transport responses discovered in our double-cross device.

**Optimizing the GANE signals via device structural engineering.** As elaborated in Figure 1, the key to realize the two-terminal detection of the out-of-plane magnetization relies on the generation of a vertical temperature gradient $\nabla_y T$ in the device. By designing a convex-shaped structure, the COMSOL simulation results confirm the appearance of two hot spots around the L-shape corners (Figure 3a), thereafter breaking the thermal uniformity along the *y*-axis in the conduction channel. Intriguingly, such an asymmetric spatial temperature distribution can be modulated by changing the device size. To quantify this size-dependent hot spot effect, the effective temperature gradient within the conduction channel of the convex-shaped device is defined as:

$$(\nabla_y T)_{\text{eff}} = \frac{1}{L_0 W_0} \left( \int_0^{W_0} \int_0^{L_0} \nabla_y T \ dxdy \right) \tag{1}$$

where the thermal-contributing channel geometry remains as ($L_0$, $W_0$) = (30 μm, 10 μm) in this work. As plotted in Figure 3b, it is observed that with the input current level of $I_0$ = 20 mA, $(\nabla_y T)_{\text{eff}}$ is linearly proportional to the length (*L*) of the protruding part with a constant width of *W* = 7.5 μm. On the contrary, under a fixed *L* = 24 μm, the amplitude of $(\nabla_y T)_{\text{eff}}$ slightly increases from 0.2 K/μm, and saturates at 0.3

K/μm when $W > 10$ μm (Supplementary Information S3). Following the optimization guidance of the COMSOL simulation results, we have accordingly fabricated the Co/Pt-based convex-shaped device with ($L_0$, $W_0$) = (24 μm, 7.5 μm), and carried out 2nd-harmonic magneto-transport measurements at room temperature. As highlighted in Figure 3c, the measured two-terminal $R_{2\omega}^{xx}$– **B** data under four different current levels of $I_0$ = 8, 12, 16 and 20 mA all exhibit distinct two-state hysteresis loops with the same clockwise polarity (*i.e.*, the combination of $(\nabla_y T)_{\text{eff}} > 0$ and $m_z > 0$ in the conduction channel corresponds to $R_{2\omega}^{xx}(+\boldsymbol{B}) < 0$ due to GANE). Moreover, the amplitude of the 2nd-harmonic signal $\Delta R_{2\omega}^{xx} = |R_{2\omega}^{xx}(+B) - R_{2\omega}^{xx}(-B)|$ is found to linearly scale with the applied current (Figure 3d), again manifesting the underlying GANE mechanism which warrants $R_{2\omega}^{\text{GANE}} = V_{2\omega}^{\text{GANE}}/I_0 = E_{2\omega}^{\text{GANE}} \cdot L_0/I_0 \propto I_0$. By further correlating the simulation and experimental results, the ANE coefficient is obtained as $S = E_{2\omega}^{\text{GANE}}/(\nabla_y T)_{\text{eff}} = 0.77$ μV/K, which is consistent with relevant data reported in [Co/Pt]$_n$-based systems[28-31]. Besides, we have also adopted a normalized parameter $\xi = \Delta R_{2\omega}^{xx}/|R_\omega^{xx} \cdot j| = 2.08 \times 10^{-13}$ A$^{-1}$·cm$^2$ (*i.e.*, where $j = I_0/(d \cdot W)$ is the current density and $d$ is the film stack thickness) to benchmark the GANE strength, and this number is comparable to the USMR coefficient (*i.e.*, which characterizes the magnetization detection efficiency) in conventional in-plane ferromagnet/heavy metal (FM/HM) heterostructures[11]. Therefore, the realization of PMA detection through the $\nabla_y T$-induced 2nd-harmonic resistance signals provides a feasible approach to electrically readout the stored information in two-terminal magnetic memories.

**Demonstration of read/write operations of the Co/Pt-based two-terminal SOT device.** To explore the GANE-enabled device applications, another [Co/Pt]$_5$ multi-layer thin film was prepared, where the Co layer thickness was fixed at 0.7 nm whereas the constituent Pt layer thicknesses varied from 1 nm to 3 nm, as illustrated in Figure 4a. This deliberately designed graded structure prevents the spins accumulated at the top and bottom channels from cancelling each other[32-34], and the resulting high SOT efficiency therefore would allow for current-induced magnetization switching (Supplementary Information S4). Considering that memory write/read operations are typically driven by DC pulses instead of sinusoid AC signals, we accordingly performed DC measurements on the fabricated convex-shaped device in this section. Specifically, with the presence of an in-plane assisted field of $B_x = +80$ mT (*i.e.*, to ensure the deterministic magnetization switching), a series of current pulses with $I_{\text{write}} = \pm 66$ mA and the pulse width of 10 ms were applied between the S/D terminals. Governed by the SOT mechanism, the charge current that conducts along the +$x$-axis (*i.e.*, $I_{\text{write}} = +66$ mA) is spin polarized towards the −$y$-direction at the Pt/Co interfaces via the spin Hall effect. On this basis, the associated spin-orbit torque $\tau_{SO} \propto \boldsymbol{m} \times \boldsymbol{\sigma} \times \boldsymbol{m}$ (*i.e.*, where $\boldsymbol{\sigma}$ is the electron spin polarization) will stabilize the magnetization along the $-z$-axis, while the reversal of the write current (*i.e.*, $I_{\text{write}} = -66$ mA) would switch the initial $-m_z$ state to the opposite direction[35-37]. After the write operation, a pair of two-terminal resistances ($R_{+I}^{xx}$, $R_{-I}^{xx}$) were recorded under positive and negative read current pulses of $I_{\text{read}} = \pm 35$ mA. Here, since $I_{\text{read}}$ would always bring about the same $(\nabla_y T)_{\text{eff}}$ in our convex-shaped device regardless of the read current direction, the sign of the DC readout signal $\Delta R_{DC}^{xx} = (R_{+I}^{xx} - R_{-I}^{xx})/2$ (*i.e.*, which is equivalent to $R_{2\omega}^{xx}$ obtained from the AC lock-in measurements) thus is exclusively determined by direction of $m_z$. In agreement the above scenario, by alternating the write current polarity, the measured $\Delta R_{xx}^{DC}$ also successively changes between $-0.81$ mΩ (red) and $+0.81$ mΩ (blue) in a highly repeatable manner (Figure 4b), therefore accomplishing the reliable electrical operations of the two-terminal SOT device based on the geometric anomalous Nernst effect. Nonetheless, it is worth mentioning that the amplitude of the $I_{\text{write}}$-enabled high-to-low contour is only 41.3% of that generated by the perpendicular magnetic field (*i.e.*, $B_z = \pm 300$ mT is large enough to magnetize the sample along the $\pm z$-axis), as displayed in Figure 4c. To elucidate this behavior, we further carried out magneto-optic Kerr effect (MOKE) microscopy on the graded Co/Pt-based device, and Figure 4d visualizes the magnetic domain structures before (upper panel) and after (lower panel) the write operation. It is clearly seen that the positive write current of $I_{\text{write}} = +66$ mA manages to achieve a SOT-driven partial magnetization switching, that is only two areas concentrated at the curved edges of the convex-shaped device (*i.e.*, where higher current densities are present) are magnetized towards the $-z$-direction. By comparing the MOKE image with the COMSOL-simulated thermal mapping, it is estimated that these two $-m_z$ magnetic domains almost coincide with the hot spot areas (33.3%), which contribute to 41.8% of the total $(\nabla_y T)_{\text{eff}}$ and serve as the main driving force to the GANE-triggered $\Delta R_{DC}^{xx}$ waveform.

**Discussion**

In conclusion, we have discovered that the non-uniform current distribution can effectively introduce in-plane temperature gradients in specially designed Co/Pt-based devices (Supplementary Information S5), and the geometric anomalous Nernst effect has led to pronounced 2$^{\text{nd}}$-harmonic magneto-transport responses whose strength is comparable to USMR-related phenomena in other FM/HM systems. By optimizing the multi-layer stack and convex-shaped device structure, efficient two-terminal write and read operations have been demonstrated in Co/Pt-based PMA SOT-device. The design principles invented in this work have bridged the fields of caloritronics and spintronics, which may also be applicable to other thermoelectric phenomena (*e.g.*, geometric spin Nernst effect, geometric spin Seebeck effect, etc.) for manipulating/probing angular momentum of spin and orbital. In addition, given that the perpendicular magnetic tunnel junction (p-MTJ) has become the mainstream choice for today's MRAM technologies due to its salient scaling capability[38-41], our work may unfold a new avenue for the development of high-density, energy-efficient memory and logic devices.

**Methods**

**Sample growth.** The [Pt(0.8 nm)/Co(0.4 nm)]$_8$/Pt(2 nm) multilayers were deposited onto 5mm × 5mm MgO(111) substrates via the DC magnetron sputtering with a base vacuum better than $5 \times 10^{-8}$ Torr, and the working argon pressure was 3 mTorr. The deposition was accomplished at room temperature with the Co and Pt deposition rates of 0.75 nm/min and 1.78 nm/min, respectively.

**Device fabrication.** Double-cross and convex-shaped devices were fabricated into μm-scale using standard photolithography and ion-beam etching methods. The etching process stopped at the insulating MgO layer to ensure that the electron current only flew through the Co/Pt multi-layer thin film, and the Cr/Au (15nm/185nm) electrodes were deposited using the e-beam evaporation to form Ohmic contacts.

**Transport Measurement.** The magneto-transport measurements were performed in an Oxford Teslatron PT system at room temperature. The AC measurements were carried out with the Keithley 6221 current sources with the current frequency of 177.78 Hz, and the first and second harmonic voltages were measured by SR830 and Zurich lock-in amplifiers, while several experimental variables such as magnetic field, and current amplitude were varied during the measurements. In DC measurements, successive positive or negative DC pulses were generated by Keithley 6221, and the output DC voltages were recorded by Keithley 2182A.

**COMSOL Simulation.** The spatial temperature and current distributions within the proposed double-cross and convex-shaped devices were simulated by COMSOL Multiphysics 5.6 based on the finite-element analysis software package, and the material parameters of each layer are listed in Supplementary Table I. During the simulation, a free triangular mesh was used in the *x-y* plane, with a minimum element size of 0.08 μm. Heat transfer between the device and the surrounding environment was set in the form of thermal convection and radiation. Here it is noted that due to the significant difference in thermal conductivity between the substrate and air, heat transfer within the device is primarily dominated by the wires and substrate. The Joule heating was introduced as the heat source and the simulated temperature distribution in each case was obtained at the equilibrium state.

**Data availability**

The data that support the findings of this study are available from the authors on reasonable request.

**Additional information**

Supplementary Information accompanies this paper.

**Acknowledgements**

This work is sponsored by the National Key R&D Program of China (grant no. 2023YFB4404000), National Natural Science Foundation of China (grant no. 92164104), and the Major Project of Shanghai Municipal Science and Technology (grant no. 2018SHZDZX02), the Shanghai Engineering Research Center of Energy Efficient and Custom AI IC, and the ShanghaiTech Material Device and Soft Matter Nano-fabrication Labs (SMN180827). X. F. Kou acknowledges support from the Zhangjiang Lab Strategic Program and the Open Fund of State Key Laboratory of Infrared Physics

**Author contributions**

X. F. Kou and C. Song supervised the study. J.M. Liu, X. F. Kou conceived and designed the experiments. J. M. Liu, H. Bai, L.Y. Liao and Y. M. Yang analyzed the experiment data. B. Rong and X.Q. Liu performed the COMSOL simulations. H. Bai, Y. H. Liu, and C. Song provided the samples. J. M. Liu, B. Rong and Y. J. Xiao performed device fabrication. J. M. Liu, B. Rong, Y. F. Zhang, Y. Z. Liang, and Q. Yao conducted the measurements. J. M. Liu and X. F. Kou wrote the manuscript. All authors discussed the results and commented on the manuscript.

**Competing financial interests**

The authors declare no competing financial interests.

## References

1. Bhatti, S. *et al.* Spintronics based random access memory: a review. *Materials Today* **20**, 530-548 (2017).
2. Chappert, C., Fert, A. & Van Dau, F. N. The emergence of spin electronics in data storage. *Nature Materials* **6**, 813-823 (2007).
3. Wolf, S. A., Lu, J., Stan, M. R., Chen, E. & Treger, D. M. The promise of nanomagnetics and spintronics for future logic and universal memory. *Proceedings of the IEEE* **98**, 2155-2168 (2010).
4. Dieny, B. *et al.* Opportunities and challenges for spintronics in the microelectronics industry. *Nature Electronics* **3**, 446-459 (2020).
5. Campbell, I., Fert, A. & Jaoul, O. The spontaneous resistivity anisotropy in Ni-based alloys. *Journal of Physics C: Solid State Physics* **3**, S95 (1970).
6. McGuire, T. & Potter, R. Anisotropic magnetoresistance in ferromagnetic 3d alloys. *IEEE Transactions on Magnetics* **11**, 1018-1038 (1975).
7. Nagaosa, N., Sinova, J., Onoda, S., MacDonald, A. H. & Ong, N. P. Anomalous hall effect. *Reviews of Modern Physics* **82**, 1539-1592 (2010).
8. Sinitsyn, N. Semiclassical theories of the anomalous Hall effect. *Journal of Physics: Condensed Matter* **20**, 023201 (2007).
9. Baibich, M. N. *et al.* Giant magnetoresistance of (001)Fe/(001)Cr magnetic superlattices. *Physical Review Letters* **61**, 2472-2475 (1988).
10. Binasch, G., Grunberg, P., Saurenbach, F. & Zinn, W. Enhanced magnetoresistance in layered magnetic structures with antiferromagnetic interlayer exchange. *Physical Review B* **39**, 4828-4830 (1989).
11. Avci, C. O. *et al.* Unidirectional spin Hall magnetoresistance in ferromagnet/normal metal bilayers. *Nature Physics* **11**, 570-575 (2015).
12. Leiva, L. *et al.* Efficient room-temperature magnetization direction detection by means of the enhanced anomalous Nernst effect in a Weyl ferromagnet. *Physical Review Materials* **6**, 064201 (2022).
13. Liu, Y.-T. *et al.* Determination of spin-orbit-torque efficiencies in heterostructures with in-plane magnetic anisotropy. *Physical Review Applied* **13**, 044032 (2020).
14. Avci, C. O. *et al.* Nonlocal detection of out-of-plane magnetization in a magnetic insulator by thermal spin drag. *Physical Review Letters* **124**, 027701 (2020).
15. Lv, Y. *et al.* Unidirectional spin-Hall and Rashba-Edelstein magnetoresistance in topological insulator-ferromagnet layer heterostructures. *Nature Communications* **9**, 111 (2018).
16. Avci, C. O., Mendil, J., Beach, G. S. D. & Gambardella, P. Origins of the Unidirectional Spin Hall Magnetoresistance in Metallic Bilayers. *Physical Review Letters* **121**, 087207 (2018).
17. Fan, Y. *et al.* Unidirectional Magneto-Resistance in Modulation-Doped Magnetic Topological Insulators. *Nano Letters* **19**, 692-698 (2019).
18. Xu, J., Phelan, W. A. & Chien, C.-L. Large anomalous Nernst effect in a van der Waals ferromagnet $Fe_3GeTe_2$. *Nano Letters* **19**, 8250-8254 (2019).

19  Sakai, A. *et al.* Giant anomalous Nernst effect and quantum-critical scaling in a ferromagnetic semimetal. *Nature Physics* **14**, 1119-1124 (2018).
20  Ikhlas, M. *et al.* Large anomalous Nernst effect at room temperature in a chiral antiferromagnet. *Nature Physics* **13**, 1085-1090 (2017).
21  Wang, S. *et al.* Controllable unidirectional magnetoresistance in ferromagnetic films with broken symmetry. *Physical Review B* **107**, 094410 (2023).
22  Rong, B. *et al.* Evidences of thermoelectrically driven unidirectional magnetoresistance from a single Weyl ferromagnet $Co_2MnGa$. *Applied Physics Letters Materials* **11** (2023).
23  Liu, J. *et al.* Observation of Moment-Dependent and Field-Driven Unidirectional Magnetoresistance in CoFeB/InSb/CdTe Heterostructures. *ACS Applied Materials & Interfaces* (2024).
24  Liu, Q., Lin, X. & Zhu, L. Absence of Spin‐Orbit Torque and Discovery of Anisotropic Planar Nernst Effect in CoFe Single Crystal. *Advanced Science* **10**, 2301409 (2023).
25  Neumann, L. & Meinert, M. Influence of the Hall-bar geometry on harmonic Hall voltage measurements of spin-orbit torques. *AIP Advances* **8** (2018).
26  Murer, C. *Thermal and electrical generation of spin currents investigated by magneto-optics*, ETH Zurich, (2020).
27  Li, J. *et al.* Field-free magnetization switching through large out-of-plane spin–orbit torque in the ferromagnetic CoPt single layers. *Applied Physics Letters* **124** (2024).
28  Lopez-Polin, G. *et al.* High-power-density energy-harvesting devices based on the anomalous Nernst effect of Co/Pt magnetic multilayers. *ACS Applied Energy Materials* **5**, 11835-11843 (2022).
29  Mochizuki, Soichiro, et al. Four-terminal sensing of laser-induced anomalous-Nernst voltages. *2023 IEEE International Magnetic Conference-Short Papers* (2023).
30  Wang, J., Miura, A., Modak, R., Takahashi, Y. K. & Uchida, K.-i. Magneto-optical design of anomalous Nernst thermopile. *Scientific Reports* **11**, 11228 (2021).
31  Frauen, A. *et al.* Magnetothermoelectric power in Co/Pt layered structures: Interface versus bulk contributions. *Physical Review B* **92**, 140402 (2015).
32  Hibino, Y., Taniguchi, T., Yakushiji, K., Kubota, H. & Yuasa, S. Observation of self-induced spin-orbit torques in Ni-Fe layers with a vertical gradient of magnetization. *Physical Review B* **109**, L180409 (2024).
33  Chen, L. *et al.* Engineering Symmetry Breaking Enables Efficient Bulk Spin-Orbit Torque-Driven Perpendicular Magnetization Switching. *Advanced Functional Materials* **34**, 2308823 (2024).
34  Zhang, K. *et al.* Efficient and controllable magnetization switching induced by intermixing-enhanced bulk spin–orbit torque in ferromagnetic multilayers. *Applied Physics Reviews* **9** (2022).
35  Song, C. *et al.* Spin-orbit torques: Materials, mechanisms, performances, and potential applications. *Progress in Materials Science* **118**, 100761 (2021).
36  Fukami, S., Anekawa, T., Zhang, C. & Ohno, H. A spin–orbit torque switching scheme with collinear magnetic easy axis and current configuration. *Nature Nanotechnology* **11**, 621-625 (2016).
37  Ryu, J., Lee, S., Lee, K. J. & Park, B. G. Current-induced spin–orbit torques for spintronic applications. *Advanced Materials* **32**, 1907148 (2020).
38  Nishimura, N. *et al.* Magnetic tunnel junction device with perpendicular magnetization films for high-density magnetic random access memory. *Journal of Applied Physics* **91**, 5246-5249 (2002).
39  Ikeda, S. *et al.* A perpendicular-anisotropy CoFeB–MgO magnetic tunnel junction. *Nature Materials* **9**, 721-724 (2010).

40  Zhang, L., Zhou, J., Li, H., Shen, L. & Feng, Y. P. Recent progress and challenges in magnetic tunnel junctions with 2D materials for spintronic applications. *Applied Physics Reviews* **8** (2021).
41  Dieny, B. & Chshiev, M. Perpendicular magnetic anisotropy at transition metal/oxide interfaces and applications. *Reviews of Modern Physics* **89**, 025008 (2017).

**Figure Captions**

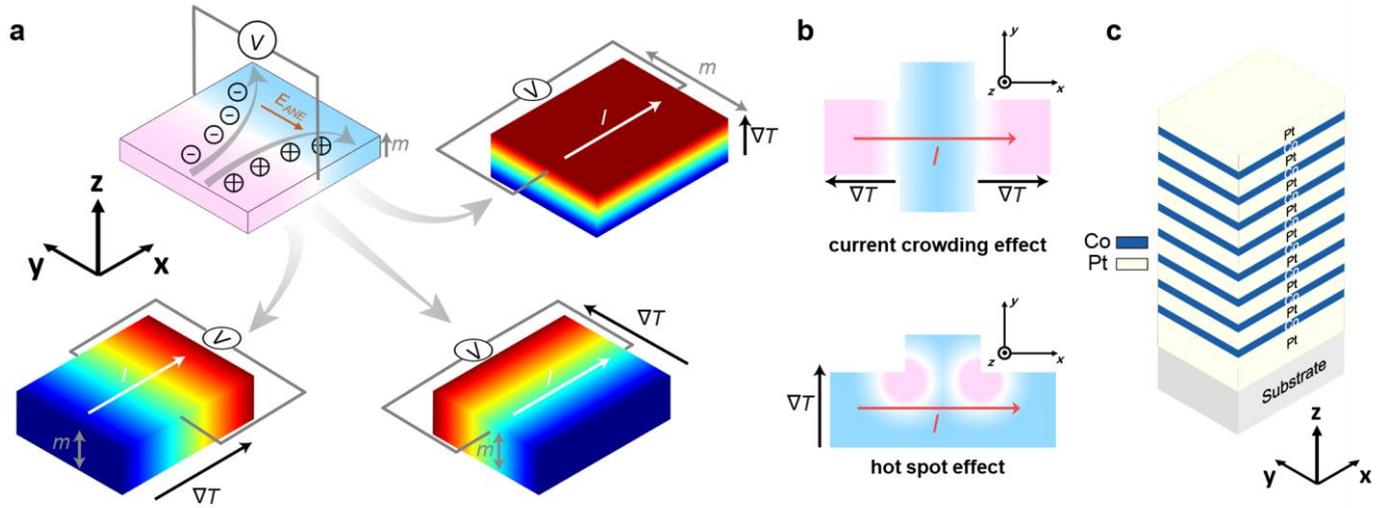

**Figure 1. Generation of in-plane temperature gradient in unconventional device structures via current crowding and hot spot effects. a.** Representation of thermoelectric Nernst effect in three cases with different ($m$, $\nabla T$) configurations. According to the relation of $\boldsymbol{E}_{\mathrm{ANE}} = S \cdot \boldsymbol{m_0} \times \nabla T$, when the perpendicular temperature gradient $\nabla_z T$ interacts with the in-plane magnetic moment $m$, a non-zero 2$^{\mathrm{nd}}$-harmonic magneto-resistance signal is generated along the current direction. In contrast, the in-plane $\nabla_y T$ and $\nabla_x T$ are able to produce $V_{2\omega}$ along the *x*-axis and *y*-axis in a PMA system. **b.** Schematics of the non-uniform temperature distributions arisen from the current crowding effect (upper panel) and hot spot effect (lower panel). **c.** [Pt(0.8 nm)/Co(0.4 nm)]$_8$ multi-layer thin film investigated in this work. A 2 nm Pt buffer layer was adopted to promote the PMA of the as-grown sample.

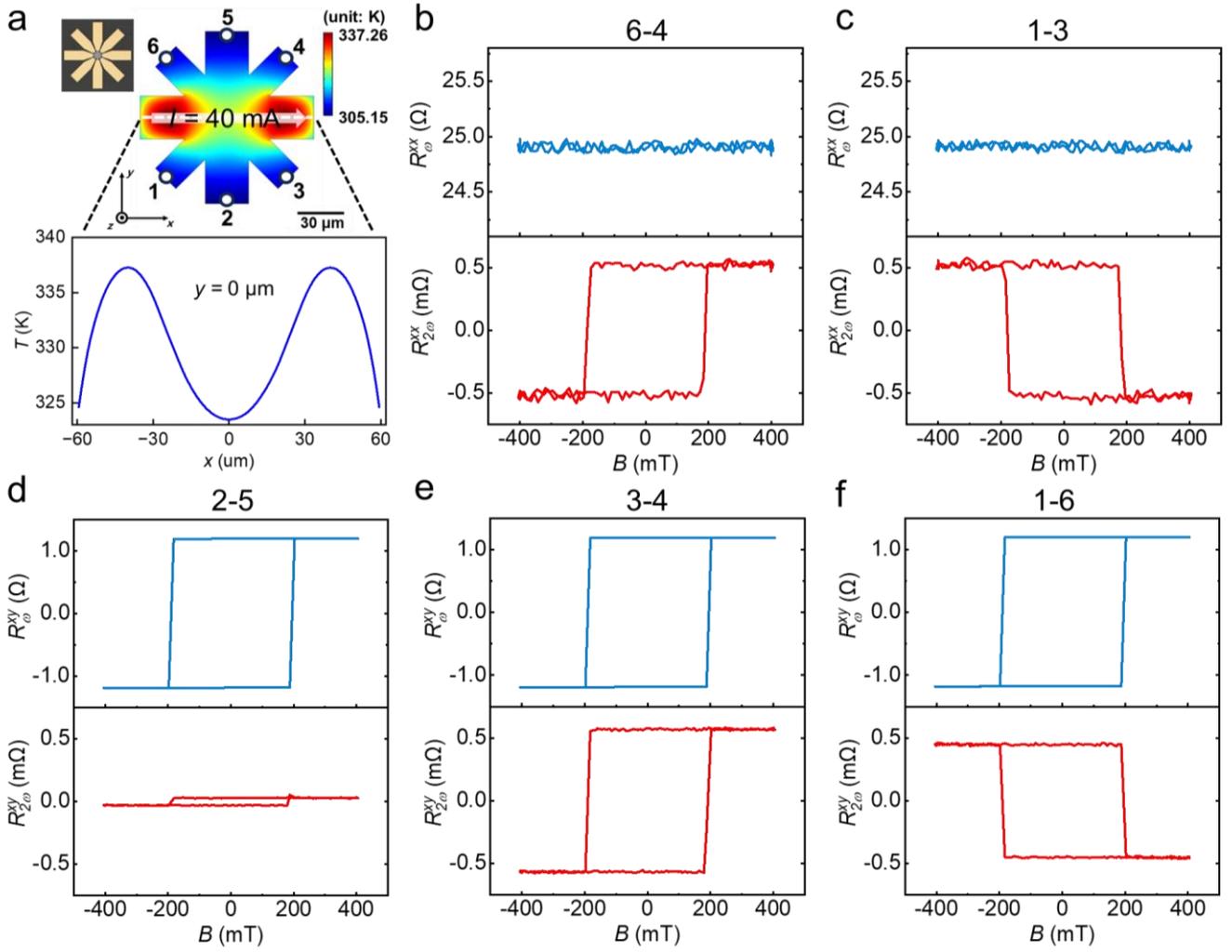

**Figure 2. Observation of the geometric anomalous Nernst effect in [Co/Pt]$_8$-based double-cross device. a.** Steady-state spatial temperature mapping of the double-cross device by COMSOL simulation. When a current of $I_0$ = 40 mA is applied through the conduction channel, the current crowding effect induces a hot-cold-hot pattern with $\Delta T$ = 15 K. **b-f.** First-order (blue) and 2$^{nd}$-harmonic (red) longitudinal and Hall resistances of $R_{\omega/2\omega}^{xx,6-4}$ (**b**), $R_{\omega/2\omega}^{xx,1-3}$ (**c**), $R_{\omega/2\omega}^{xy,2-5}$ (**d**), $R_{\omega/2\omega}^{xy,3-4}$ (**e**), and $R_{\omega/2\omega}^{xy,1-6}$ (**f**) of the fabricated device. The square-shaped $R_{\omega}^{xy,2-5}$, $R_{\omega}^{xy,3-4}$, and $R_{\omega}^{xy,1-6}$ hysteresis loops confirm the robust perpendicular magnetic anisotropy of the [Co/Pt]$_8$ thin film, while the opposite polarities of the 2$^{nd}$-harmonic magneto-transport signals manifest the underlying GANE origin.

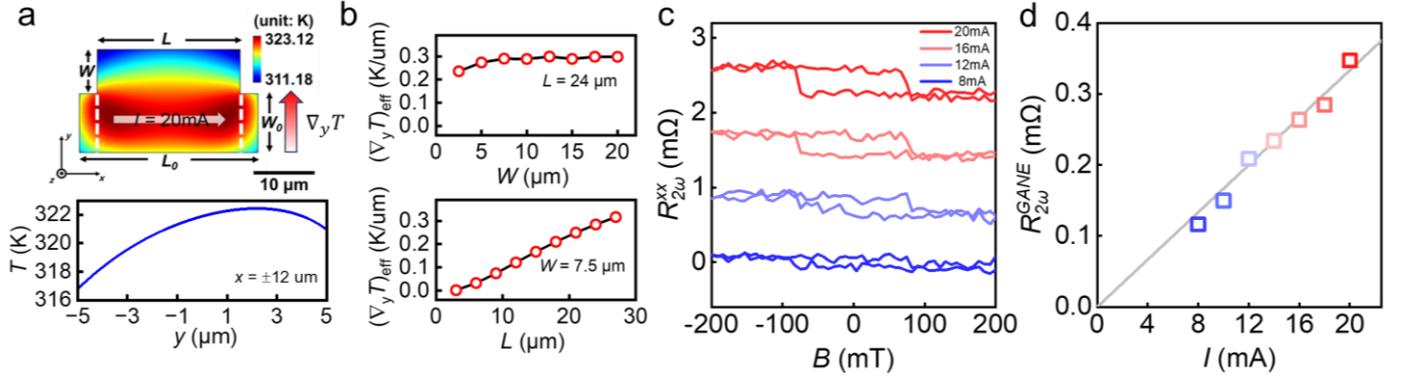

**Figure 3. Optimization of the GANE effect in the [Co/Pt]$_8$ convex-shaped device. a.** Steady-state spatial temperature mapping of the convex-shaped device by COMSOL simulation. When a current of $I_0$ = 20 mA is applied, the hot spot effect gives rise to a non-zero temperature gradient along the *y*-direction. Here, ($L_0$, $W_0$) = (30 μm, 10 μm) is the geometry of the conduction channel, while the variables (*L*, *W*) = (24 μm, 7.5 μm) refer to the size of the protruding part. The hot spot effect induces a thermal gradient of $(\nabla_y T)_{\text{eff}}$ = 0.28 K/μm. **b.** Evolution of the effective temperature gradient $(\nabla_y T)_{\text{eff}}$ as functions of *L* and *W* by COMSOL simulations. It is seen that $(\nabla_y T)_{\text{eff}}$ scales linearly with *L*, whereas it becomes saturated when *W* ≥ 7.5 μm. **c.** Measured two-state $R_{2\omega}^{xx} - B$ hysteresis loops of the fabricated (*L*, *W*) = (24 μm, 7.5 μm) device under four write currents of $I_0$ = 8 mA, 12 mA, 16 mA, and 20 mA, respectively. **d.** The amplitude of the GANE-induced 2$^{\text{nd}}$-harmonic magneto-resistance $\Delta R_{2\omega}^{xx}$ is linearly proportional to the applied current level.

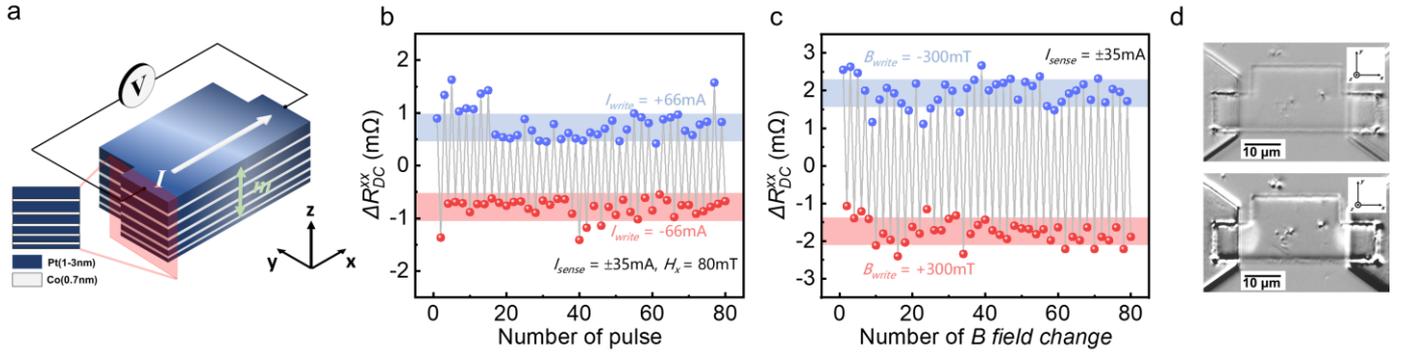

**Figure 4. Demonstration of read and write operations of the graded [Co/Pt]$_5$-based two-terminal SOT device. a.** Schematic of device structure. All the five Co layers have a fixed 0.7 nm thickness, while the Pt layer thickness gradually increases from 1 to 3 nm so that the overall SOT efficiency is maximized. **b-c.** Repetition DC measurements of the S/D resistance after 40 cycles of write-and-read operations with the presence of an in-plane assisted magnetic field of $B_x$ = 80 mT. The write and read currents are set as $I_{\text{write}}$ = ±66 mA and $I_{\text{read}}$ = ±35 mA, respectively. In each cycle, the magnetization was switched by alternating the current (**b**) and magnetic field (**c**) polarities. The square-waveforms are in phase in both measurements, yet the amplitude of current-driven $\Delta R_{\text{DC}}^{xx}$ is 41.3% compared to the magnetic field-dependent data. **d.** Supplementary MOKE images before (upper panel) and after (lower panel) the write operation. It is seen that only the hot spots at the curved edges of the convex-shaped device are switched.